\documentclass[11pt,aps,prd,superscriptaddress,notitlepage]{revtex4-1}
\usepackage{amsmath}
\usepackage[dvips]{graphicx}
\usepackage[english]{babel}
\newcommand{\be}{\begin{equation}}
\newcommand{\ee}{\end{equation}}
\newcommand{\ben}{\begin{eqnarray}}
\newcommand{\een}{\end{eqnarray}}
\newcommand{\no}{\noindent}

\begin{document}
\footnotesize
\title{Holographic dark energy described at the Hubble length}
\author{Ivan Duran\email{ivan.duran@uab.cat}}
\affiliation{Departament de F\'{i}sica, Universitat Aut\`{o}noma de Barcelona Bellaterra, Spain}
\author{Luca Parisi\email{parisi@sa.infn.it}}
\affiliation{Dipartimento di Fisica ``E.R.Caianiello", Universit\`{a} di Salerno and\\
 INFN, Sezione di Napoli, GC di Salerno, Fisciano (Sa), Italy}

\

\begin{abstract}
We consider holographic cosmological models of dark energy in which the infrared cutoff is set by the Hubble's radius. We show that any interacting dark energy model with a matter like term able to alleviate the coincidence problem (i.e., with a positive interaction term, regardless of its detailed form) can be recast as a noninteracting model in which the holographic parameter $c^{2}$ evolves slowly with time. Two specific cases are analyzed. First, the interacting model presented in \cite{PavonDuranZimdahl} is considered, and its corresponding noninteracting version found. Then, a new noninteracting model, with a specific $c^{2}(z)$ expression, is proposed and analyzed along with its corresponding interacting version. We constrain the parameters of both models using observational data, and show that they can be told apart at the perturbative level.
\end{abstract}
\maketitle

\section{Introduction}\label{sec:Intro}

Nowadays the Universe appears to be undergoing a phase of accelerated expansion \cite {PerlmRiess}. This can be explained through modified gravity theories \cite{modGrav} and by general relativity (GR) modulo introducing an exotic component, dubbed   dark energy (DE). We only know for certain that DE is endowed with a hugely negative pressure (of the order of its energy density) and that it should be distributed rather evenly across space; see \cite{recentreviews} to learn about the state of the art. Considering a spatially flat Friedman-Robertson-Walker (FRW) metric in the frame of GR the Friedmann equations read

\ben\label{eq:FE1}
H^2&=&\frac{M_{P}^{-2}}{3} \left(\rho_{M}+\rho_{X}\right) \, ,\\
\label{eq:FE2}
\dot{H}&=&-\frac{M_{P}^{-2}}{2} \left( \rho_{M}+\rho_{X}+P \right) \, ,
\een
where $M_{P} = (8 \pi G)^{-1/2}$ is the reduced Planck mass, $P=w \rho_{X}$, and $\rho_{X}$ and $\rho_{M}$ indicate the DE and dark matter (DM) energy densities, respectively.

Thus far, the most successful candidate for DE is the cosmological constant $\Lambda$, which together with cold dark matter and radiation form the standard cosmological model, $\Lambda$CDM. Regrettably this model suffers from two main problems: $(i)$ the cosmological constant value, $\Lambda$,  is many orders of magnitude lower than that expected for the vacuum energy density from quantum field theory, and $(ii)$ the coincidence problem.\\  The latter can be alleviated by introducing a suitable interaction between DE and DM, which appears consistent with observations from relaxed galaxy clusters \cite{He:2010ta,Abdalla, Tarrant}. For interacting models, the conservation equations are

\be\label{eq:evolEqIntMX}
\dot{\rho}_{M} + 3H \rho_{M} = Q    \qquad \text{and}\qquad
\dot{\rho}_{X} + 3H (1+w)\rho_{X} = - Q  \, .
\ee
\no where $Q$ is the interaction term. \\
Whatever the nature of DE it seems reasonable that it respects
the holographic principle. The latter asserts that the number of
relevant degrees of freedom of a system dominated by gravity must
vary as the area of the surface bounding the system
\cite{gerard-leonard}. In addition, the energy density of any
given region should be bounded by that ascribed to a Schwarzschild
black hole that fills the same volume \cite{cohen}. Mathematically
this condition reads $\rho_{X}\leq M_{P}^{2}\, L^{-2}$, where
$L$ stands for the size of the considered region (i.e., the infrared (IR) cutoff). This expression is most
frequently written in its saturated form
\be \rho_{X}= \frac{3 M_{P}^{2}\,  c^{2}}{L^{2}}\, . \label{eq:rhox}
\ee
Here $c^{2}$ is a dimensionless parameter -very often, assumed
constant- that summarizes the uncertainties of the theory (such as
the number particle species and so on); the factor $3$ was
introduced just for mathematical convenience.

When dealing with holographic DE one must first specify the
IR cutoff. With the lack of clear guidance different
expressions have been adopted. The most relevant ones are the
Hubble radius, $H^{-1}$, see e.g. \cite{hubbleradius}
, the future events horizon $R_{h}=\int^{\infty}_{t}\frac{dt}{a}$ \cite{MLi}, that suffers from a severe circularity problem, and the Ricci length, $R_{CC}$, i.e., $L =R_{CC}= (\dot{H} +2H^{2})^{-1/2}$ -see
e.g. \cite{gao,lxu,suwa,DuranPavonPrd,lepe}. The rationale behind the latter is
that it corresponds to the size of the maximal perturbation
leading to the formation of a black hole \cite{brustein}.

It has been argued that an IR cutoff defined by the Hubble radius, $L=H^{-1}$, can not lead to an accelerated Universe, since DE evolves as $H^{2}$, that is, as pressureless matter, therefore producing a decelerated expansion \cite{MLi}. However, if DM and DE interact via Eqs. (\ref{eq:evolEqIntMX}), with $Q>0$, the present epoch of accelerated expansion can be realized \cite{ZimdahlPavon}.

Differentiating the expression $\rho_{X}=3M_{P}^{2}c^{2}H^{2} $, that describes DE density for Hubble holographic models, and using Eqs. (\ref{eq:FE2}) and (\ref{eq:evolEqIntMX}.2), the equation of state (EoS) parameter
\be\label{eq:wi}
w=-\frac{Q}{3(1-c^{2})\rho_{X}H}
\ee
\no follows. Note that $Q$ must be positive, otherwise $w$ would be positive, and the Universe would not accelerate. Further, if $Q$ were negative, the second principle of thermodynamics would be violated \cite{PavonWang}. A comprehensive study of holographic DE models can be found in \cite{DelCampo}.

An alternative approach to simultaneously solve the coincidence problem and describe the late time acceleration was taken in \cite{Pavon:2005yx}. To get an evolving energy density ratio, $r\equiv \rho_{M}/\rho_{X}$, with the Hubble scale as IR cutoff, the $c^2$ parameter was promoted to vary with time, though slowly, in such a way that $\left(c^{2}\right)\dot{}\geq 0$. In that scenario, the holographic bound is progressively saturated in the sense that $c^{2}_{t\rightarrow\infty}=$ constant $>0$.

Likewise, it was shown that a variable $c^2$ significantly alleviates the coincidence problem also for cosmological model with nonvanishing spatial curvature \cite{Pavon:2006qm}. An extended analysis of nonsaturated holographic DE models with nonvanishing spatial curvature was recently presented in \cite{Cardenas:2010wx}. There, it was shown that a variable $c^2$ is compatible with current precision data, which favor a small $\Omega _k$, but high enough to have significant cosmological consequences.

Moreover, a nonsaturated holographic bound provides a transition from a decelerated to an accelerated era not only for standard holographic DE models, but also for a class of generalized holographic DE models in which the gravitational coupling is promoted to a time dependent quantity \cite{Guberina:2006qh}.

In \cite{Xu:2009ys} the role played by a variable $c^2$ parameter in a model with the IR cutoff set by the Hubble's length was discussed in connection with the behavior of the effective EoS of time variable cosmological constant models.  A parametrized expression for $c^2$ was proposed and tested using observational data. Notice, however, that this parametrization of $c^{2}$, leads to a negative DM density in the future.

In \cite{Wei:2009au} a modified holographic DE model with a variable $c^{2}$ was considered in which the IR cutoff was set by the Ricci scale, $L = R_{CC}$. Three different parametrizations were proposed and tested with cosmological data leading to consistent results.

Recently, a comprehensive analysis of variable $c^2$ holographic DE models for the three widely used infrared cutoff scales, namely the Hubble's length, the Ricci's length and the particle horizon length, was presented in \cite{RadicellaPavon}. It was argued that the $c^2$ term appearing in the conventional formula for holographic DE should not be assumed constant in general, except for the particular case of DE with the Ricci's length as infrared cutoff.

In what follows, quantities referring to holographic models with variable $c^{2}$ will be noted by a tilde. They obey
\be\label{eq:rhoMX}
\tilde{\rho}_{M}=3M_{P}^{2}(1-\tilde{c}^{2})H^{2}  \qquad \text{and}\qquad
\tilde{\rho}_{X}=3M_{P}^{2}\tilde{c}^{2}H^{2} \, .
\ee
as well as
\be\label{eq:EvolutionMX}
\dot{\tilde{\rho}}_{M}=-3H\tilde{\rho}_{M} \qquad \text{and}\qquad
\dot{\tilde{\rho}}_{X}=-3H(1+\tilde{w})\tilde{\rho}_{X} \,,
\ee
\no i.e., we assume that their energy densities conserve separately.\\
In this paper, we study some general features of holographic DE models where the
IR cutoff is defined by the Hubble radius. By considering both points of view,
we shall show that, in general, identical cosmological backgrounds can be described by an interacting holographic DE model where the holographic parameter $c^{2}$ is a constant or, alternatively, by a noninteracting holographic DE model in which the $\tilde{c}^{2}$ parameter depends weakly on time. In spite of the global evolution in both scenarios being the same, the energy densities, the EoS parameters and so on, can behave rather differently.

We remark that in any holographic interacting model defined at the Hubble scale, independently of the nature of the interaction, DE and DM share the same dependence  on $H$, and thus present the same background evolution. By rewriting Eqs. (\ref{eq:evolEqIntMX}) as
\ben\label{eq:evolEqNoIntMX}
\dot{\rho}_{M} + 3H \left(1-\frac{Q}{3H\rho_{M}}\right)\rho_{M} = 0 \qquad
\text{and} \qquad
\dot{\rho}_{X} + 3H \left(1+w+\frac{Q}{3H\rho_{X}}\right)\rho_{X} = 0
\een
and using Eq.(\ref{eq:wi}) both, DM and DE, are seen to share the same effective
EoS parameter, $w_{eff}\equiv-\frac{Q}{3H\rho_{M}}$. We will return to this later on.

We shall also r-consider the coincidence problem from the viewpoint of noninteracting models where the energy densities ratio, $\tilde{r}\equiv \tilde{\rho}_M / \tilde{\rho}_X $, is not a constant.

The plan of the paper is as follows. In Sec. \ref{Equivalence}, we show the
equivalence at the background level between interacting holographic DE
models and noninteracting holographic DE models in which the holographic
parameter $\tilde{c}^{2}$ depends on time. In Sec. \ref{Constraints},
we briefly introduce the cosmological tools (e.g. luminosity distance,
angular distance, and so on) used to characterize the models studied in
this work. In Sec. \ref{sec:EX1}, a specific model is considered,
namely, the interacting DE model of Ref. \cite{PavonDuranZimdahl}.
Its corresponding non-interacting version is derived and their
Lagrangian formulation is presented. In Sec. \ref{sec:EX2}, another model, in which a novel expression for $\tilde{c}^{2}$ is proposed, aimed to alternatively describe the late behavior of both DE and DM energy densities, is presented and contrasted with the concordance $\Lambda$CDM model. In section \ref{Perturbations}, we show that the background
equivalence is broken at the perturbative level. Finally Sec.
\ref{Conclusions} summarizes our main results.

\section{Background equivalence between interacting and $\tilde{c}^{2}$ models}
\label{Equivalence}

To show that every interacting model can be considered as a noninteracting one (with a $\tilde{c}^{2}$ parameter varying in time) at the background level (i.e., both interpretations share the same Hubble function), we must first verify that both DE and DM energy densities are positive. By Eqs. (\ref{eq:rhoMX}), this condition implies $0 \leq \tilde{c}^{2} \leq 1$.

From Eqs. (\ref{eq:FE1}), (\ref{eq:FE2}) and (\ref{eq:rhoMX}.2) we obtain
\be\label{eq:Friedmann2}
\dot{H}=-\frac{3}{2}H^{2}\left(1+\tilde{c}^{2}\tilde{w}\right) \,.
\ee
For the noninteracting model, from Eq. (\ref{eq:EvolutionMX}.1) we have that $\tilde{\rho}_{M}=\tilde{\rho}_{M\,0}a^{-3}$ remains always positive.

We must also verify that $\tilde{\rho}_{X}$ is positive, i.e.,  that $\tilde{c}^{2}\geq 0$. Since by hypothesis interacting and $\tilde{c}^2$ models share the same $H(z)$, Eq. (\ref{eq:Friedmann2}) implies
\be\label{eq:RelCsWs}
\frac{\tilde{c}^{2}(z)}{c^{2}}=\frac{w(z)}{\tilde{w}_(z)}\,,
\ee
\no where $z$ is the redshift ($z=\frac{1}{a}-1$). Thus, if the noninteracting DE has a negative EoS parameter $\tilde{w}(z)$, since $w(z)\,\leq\, 0$ and $c^{2}\,\geq\, 0$, then $\tilde{\rho}_{X}$ and $\tilde{c}^{2}(z)$ will always be positive.\\\\
Since $\rho_{M}\geq 0$ and $\tilde{c}^{2} \leq 1$, by differentiating Eq. (\ref{eq:rhoMX}.2), using  Eqs. (\ref{eq:wi}), (\ref{eq:Friedmann2}) and (\ref{eq:RelCsWs}), and recalling that $Q>0$, we obtain
\be\label{eq:cPositive}
(\tilde{c}^{2})\dot{}=\frac{1-\tilde{c}^{2}}{\rho_{M}}Q\, \geq 0\,.
\ee
\no As shown in \cite{RadicellaPavon}, this condition must be fulfilled for thermodynamical reasons.  Thus, we conclude that the Hubble function of any holographic interacting model with $Q\geq0$ and $c^{2}$ constant also corresponds to a holographic noninteracting model ($Q=0$) of DE ($\tilde{w}\leq 0$), with $\tilde{c}^{2}$ obeying $0 \leq \tilde{c}^{2}(t)\leq 1$ and $\left( \tilde{c}^{2} \right)\dot{}\geq0$, and vice versa.
It is also noteworthy that despite both models being equivalent at the background level, they share the same $H(z)$ and their energy components evolve diversely. In the interacting case, the Hubble function will never have $c^{2}$ as a free parameter, but it will be multiplied by the constants in the interacting term $Q$, as can be seen by using Eq. (\ref{eq:FE2}) in Eq. (\ref{eq:wi}). This means that, as long as $\Omega_{X}=c^{2}$, the $\Omega_{X\,0}$ parameter cannot be fitted because we have neither observational nor theoretical constraints on the interaction, and so the dependence on them is degenerated. On the other hand, in the $\tilde{c}^{2}$ model it has the fixed value, $\tilde{\Omega}_{X\, 0}=\tilde{c}^{2}(z=0)$.
Consequently, at background level it is not possible, in principle, discriminate interacting models from $\tilde{c}^{2}$ ones. Nevertheless, as we shall see in Sec. \ref{Perturbations}, they are distinguishable at the perturbative level.
\section {Observational constraints}\label{Constraints}
To constrain the free parameters of the Hubble holographic models presented below (Sec. \ref{sec:EX1} and  \ref{sec:EX2}), we shall use observational data from SN Ia Union2 set (557 data points) \cite{Amanullah}, BAO \cite{Percival}, the acoustic scale $l_{A}$ \cite{Komatsu}, gas mass fractions in galaxy clusters as
inferred from x-ray data (42 data points) \cite{Allen} and the Hubble rate (15
data points) \cite{Riess-2009,gazta,simon,stern}. As usual, the
likelihood function is defined by $ {\cal L} \propto \exp (-
\chi^{2}/2)$. The best fit parameter values can be found by minimizing the sum $
\chi^{2}_{\rm total} = \chi^{2}_{SN} \, + \,
\chi^{2}_{bao} \,+ \,
\chi^{2}_{l_{A}} \, + \, \chi^{2}_{x-rays}+ \, \chi^{2}_{Hubble}$.
\subsection {SN Ia}
We contrast the theoretical distance modulus
\be \mu_{th}(z_{i}) = 5\log_{10}\left(\frac{d_{L}}{{10{\rm
pc}}}\right)\, + \,\mu_{0}\, , \label{modulus} \ee
where $\mu_{0} = 42.38 \, -\, 5\log_{10} h$, with the observed
distance modulus $\mu_{obs}(z_{i})$ of the 557 SN Ia compiled in
the Union2 set \cite{Amanullah}. Here $h\equiv\frac{H_{0}}{100}$ with $H_{0}$ in $km/s/Mpc$. The latter assembly is much
richer than previous SN Ia compilations and has some other
advantages, especially the refitting of all light curves with the
SALT2 fitter and an enhanced control of systematic errors. In
(\ref{modulus}) $d_{L} = (1+z) \int_{0}^{z}{\frac{dz'}{E(z';{\bf
p})}}$ denotes the Hubble-free luminosity distance, with ${\bf p}$
the model parameters. To eliminate the effect of the nuisance parameter $\mu_{0}$ we
resort to the method of \cite{Nesseris-Perilovorapoulos_0}.
\subsection{BAO}
Pressure waves originated from cosmological perturbations in the
primeval baryon-photon plasma produced acoustic oscillations in
the baryonic fluid. These oscillations have been unveiled by a
clear peak in the large scale correlation function measured from
the luminous red galaxies sample of the Sloan Digital Sky Survey
(SDSS)  at $z = 0.35$ \cite{Eisenstein} as well as in the Two
Degree Field Galaxy Redshift Survey (2dFGRS) at $z = 0.2$
\cite{Percival}. These peaks can be traced to expanding spherical
waves of baryonic perturbations with a characteristic distance
scale
\begin{equation}\label{eq:BAO}
D_{v}(z_{BAO})=\left[\frac{z_{BAO}}{H(z_{BAO})}\,d_{A}(z_{BAO})^{2}\right]^{\frac{1}{3}}
\, ,
\end{equation}
\no where $d_{A}(z)=(1+z)^{-1}\int^{z}_{0}\frac{dz'}{H(z')}$. Data from SDSS  and 2dFGRS observations yield
$D_{v}(0.35)/D_{v}(0.2)=1.736 \pm 0.065$, a nearly model independent value \cite{Percival}. The most constraining values $r_{s}(z_{\star})/D_{v}(z_{i})$, for $z_{i}=0.2, 0.35, 0.278$, obtained in \cite{Percival,Kazin}, are not model independent but obtained just for flat and open $\Lambda$CDM and $w$CDM models.
\\The sound horizon radius is given by
\be\label{soundHorizon}
r_{s}(z)=\int^{\infty}_{z}\frac{c_{s}}{H(z')}dz'
\ee
\no and the sound speed before decoupling is $c_{s}=\frac{1}{\sqrt{3}}\left(1+\frac{3\Omega_{B}}{4\Omega_{\gamma}}\right)^{-\frac{1}{2}} $, where $\Omega_{B}$  and $\Omega_{\gamma}$ are the fractional density parameters of baryons and photons, respectively.
\subsection{Acoustic scale $l_{A}$}
The cosmic microwave background (CMB) power spectrum is sensitive to the distance
to the decoupling epoch via the peaks location. More specifically, with CMB data one can measure two
distance ratios. The first one is the acoustic scale, $l_{A}$, described by the angular distance to the decoupling surface divided by the sound horizon radius at that time
\begin{equation}\label{eq:acousticScale}
l_{A}=\pi\frac{d_{A}(z_{\star})}{r_{s}(z_{\star})} \, ,
\end{equation}
\no where $d_{A}(z)$ and $r_{s}(z)$ are comoving quantities.
\\The other one is the CMB shift parameter, $R$, defined by the ratio of the distance to decoupling
to the Hubble horizon, $H^{-1}(z_{\star})$, at that time
\begin{equation}\label{eq:cmbShift}
R=\sqrt{\Omega_{M\,0}}H_{0}\int^{z_{\star}}_{0}\frac{dz}{H(z)} \, .
\end{equation}
The latter expression assumes a negligible DE density at decoupling, and should be modified when used in models other than $\Lambda$CDM or $w$CDM \cite{Komatsu, Kowalski, Sollerman}. For this reason (our interacting model has a constant DE-DM ratio), we will just use the acoustic scale , $l_{A}$, to constrain the models. However, to describe the Universe before decoupling one should include also the baryon and radiation energy component (see Appendix \ref{app:Rad}).
\subsection {Gas mass fraction}\label{subSec:GMF}
As is well known, a very useful indicator of the overall cosmic
ratio $\Omega_{baryons}/\Omega_{M}$, nearly independent of
redshift,  is the fraction of baryons in galaxy clusters, $f_{gas}$; see
\cite{White}. This quantity can be determined from the x-ray flux
originated in hot clouds of baryons, and it is related to the
cosmological parameters by $f_{gas} \propto d_{A}^{3/2}$, with
$d_{A}$ the angular diameter distance to the cluster.

We used measurements by the Chandra satellite of 42 dynamically
relaxed galaxy clusters in the redshift interval $0.05 < z < 1.1$
\cite{Allen}. In  fitting the data we resorted to the empirical
formula
\begin{equation}\label{eq:Allen}
f_{gas}(z)=\frac{K \, A \, \gamma\, b(z)}{1\, + \, s(z)}
\frac{\Omega_{B0}}{\Omega_{M0}}\left(\frac{d_{A}^{\Lambda
CDM}}{d_{A}}\right)^{3/2}
\end{equation}
(see Eq. (3) in Ref. \cite{Allen}) in which the $\Lambda$CDM model
serves as a fiducial model. Here, the parameters $K$, $A$, $\gamma$,
$b(z)$ and $s(z)$ model the abundance of gas in the clusters. We
set these parameters to their respective best fit values for the
considered model, but in the intervals defined in \cite{Allen}.
\subsection {History of the Hubble function}
Recently, high precision measurements by Riess {\em et al.} at $z
= 0$, from the observation of 240 Cepheid variables of rather
similar periods and metallicities \cite{Riess-2009}, as well as
measurements by Gazta\~{n}aga {\em et al.}, at $z = 0.24, \, 0.34,
{\rm and}\, 0.43 \,$ \cite{gazta}, who used the BAO peak position
as a standard ruler in the radial direction, have somewhat
improved our knowledge about $H(z)$. However, at redshifts above,
say, $0.5$ this function remains largely undetermined. Yet, to constrain the holographic model we have considered these
four data alongside 11 noisier data in the redshift interval $0.1
\lesssim z \lesssim 1.8$, from Simon {\em et al.} \cite{simon} and
Stern {\em et al.} \cite{stern}, obtained from the differential
ages of passive-evolving galaxies and archival data.
\section{Model 1}\label{sec:EX1}
\begin{figure*}[!htb]
  \begin{center}
    \begin{tabular}{c}
      \resizebox{90mm}{!}{\includegraphics{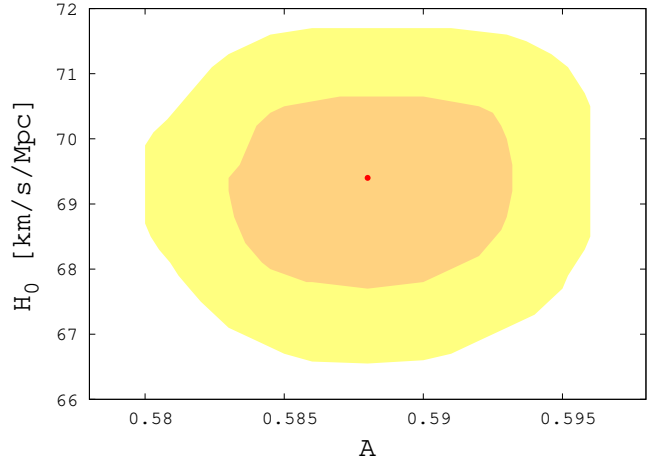}}\\
    \end{tabular}
    \caption{\scriptsize 1$\sigma$ and 2$\sigma$ confidence regions for the parameters $A$ and $H_{0}$ of model 1. The plot holds for both scenarios, the interacting and the $\tilde{c}^{2}(t)$ (noninteracting). The dot indicates the best fit values.}
    \label{fig:elipsesEx1}
 \end{center}
\end{figure*}

\begin{figure*}[!htb]
  \begin{center}
    \begin{tabular}{cc}
      \resizebox{80mm}{!}{\includegraphics{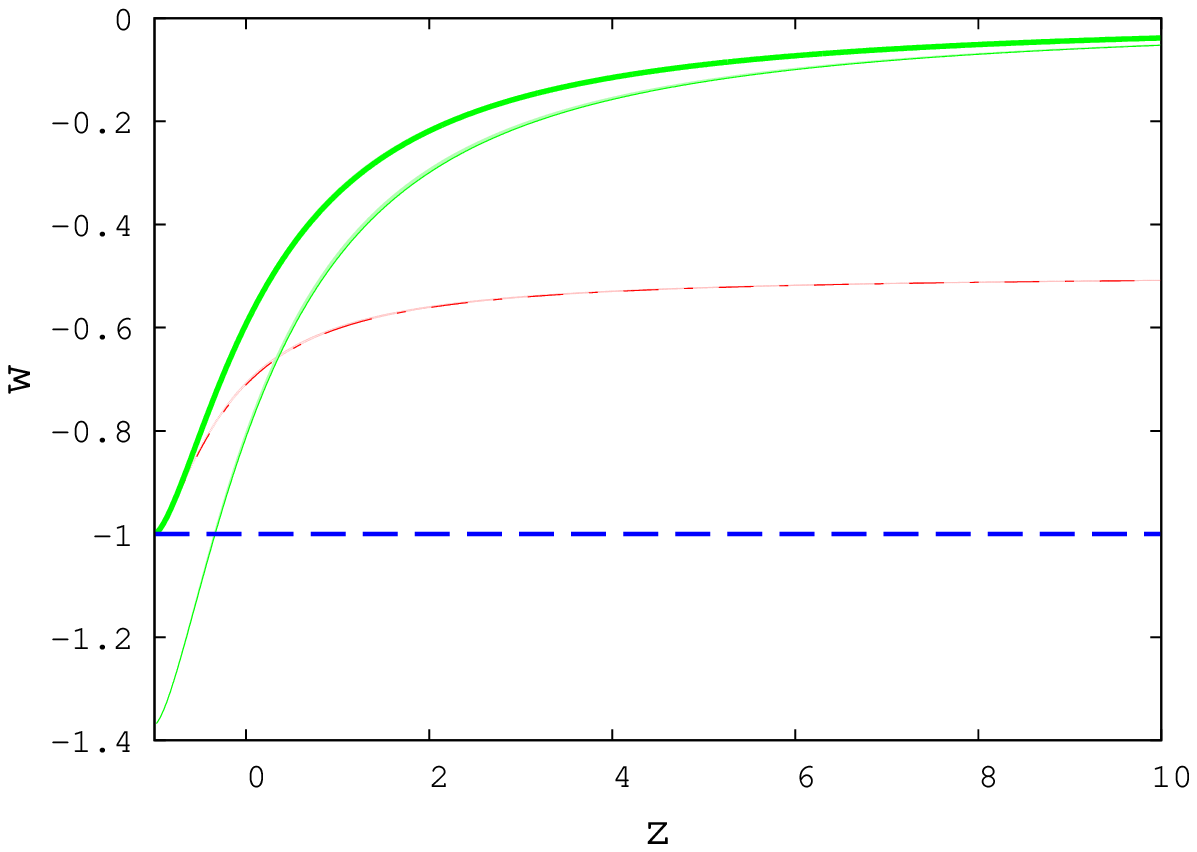}}&
      \resizebox{80mm}{!}{\includegraphics{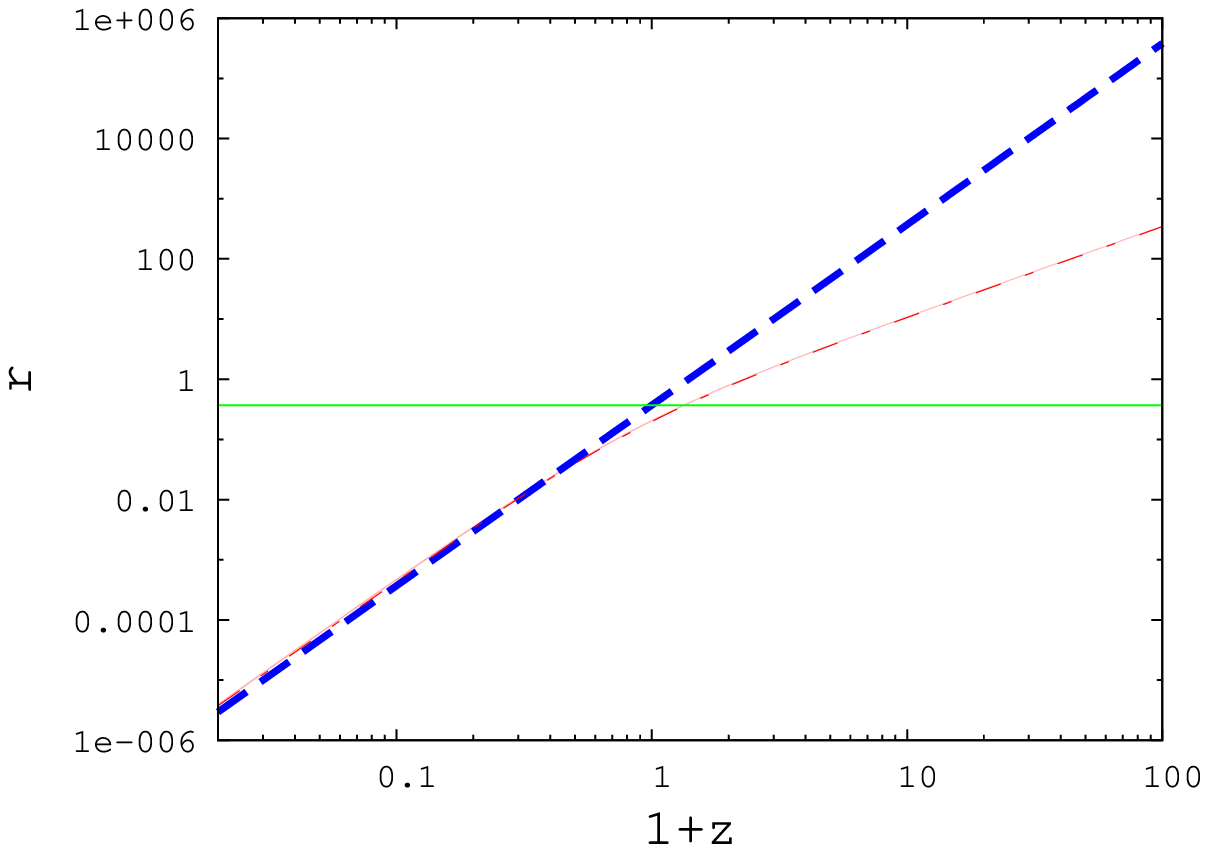}}\\
    \end{tabular}
    \caption{\scriptsize Left panel: EoS parameter for the interacting ($w$ thin line, and $w_{eff}$ thick line), the $\tilde{c}^{2}$  and the $\Lambda$CDM models. Right panel: energy densities ratios, $r\equiv\rho_{M}/\rho_{X}$, versus $1+z$ for the $\Lambda$CDM, the interacting and the $\tilde{c}^{2}$ models. All the graphs were plotted using the best fit values of the parameters, shown in table \ref{tab:ex1a}. Solid (green) lines are used for the interacting case, thin dot-dashed (red) lines for the $\tilde{c}^{2}$ model, and thick short dashed (blue) for $\Lambda$CDM. The 1$\sigma$ region of the parameters is also plotted, but due to the very small errors, it results are nearly inappreciable.}
    \label{fig:ex1}
 \end{center}
\end{figure*}

Let us now consider the holographic interacting model studied in \cite {PavonDuranZimdahl} to construct its equivalent $\tilde{c}^{2}(t)$ model. In the former the IR cutoff is also set by the Hubble length and the interaction term was taken as $Q \equiv 3AH_{0}\rho_{M}$, with $A$ a semipositive definite constant, related to the constant decay rate, $\Gamma$, of DE into DM by $A\equiv\frac{\Gamma}{3H_{0}r}$. The corresponding Hubble function is
\be\label{eq:H1}
H=H_{0}\left(A+(1-A)(1+z)^{\frac{3}{2}}\right)\, ,
\ee
\no and the equation of state parameter
\be\label{eq:wInt}
w = -\frac{A}{\Omega_{X}}\frac{H_{0}}{H},
\ee
follows from Eq. (\ref{eq:wi}) and the fact that $\Omega_{X}=c^{2}$ and $\Omega_{M}=1-c^{2}$.
Notice that here the density parameters, $\Omega_{i}\equiv\frac{\rho_{i}}{3M_{P}^{2}H^{2}}$ ($i=M,X$) are constant. The model fits reasonably well the observational data and is consistent with the age of the old quasar APM 08279+5255 \cite{Quasar}, something that $\Lambda$CDM is not at $1\sigma$ confidence level.\\
This model presents an unexpected similarity to the Chaplygin gas model \cite{Bento}. To see this, bear in mind that in any Hubble holographic interacting model DM and DE share the same effective EoS parameter, $w_{eff}=-Q/(3H\rho_{M})$. Multiplying it by the total energy density we get the total pressure
\be\label{eq:PInt}
P = w_{eff}\, \rho = -3^{\frac{1}{2}}A\,M_{P}\,H_{0}\,\rho^{\frac{1}{2}} \,,
\ee
\no with $A$ a semi-positive definite constant. This expression is formally identical to the pressure of the generalized Chaplygin gas, $P=-\beta\rho^{-\alpha}$ \cite{Bento}. As is well known, for $\alpha<0$ and $\beta>0$, it may imply instabilities since the squared adiabatic sound speed ($c_{s\,a}^{2}\equiv\frac{\dot{P}}{\dot{\rho}}$) is negative . Becaue of the interaction, we can take account of non adiabatic processes, and so consider an effective speed of sound. In Appendix \ref{app:Lag} we find a Lagrangian formulation with a standard scalar field $\phi$ for both models (the interacting and the $\tilde{c}^{2}$) with an effective speed of sound given by $c_{s}^{2}=1$. \\
Notice that while in the interacting case $r\equiv\rho_{M}/\rho_{X}$ is a constant, in the noninteracting one, $\tilde{r}$ may vary with time. We expand $H^{2}(z)$ and assume that the term proportional to $(1+z)^{3}$, corresponds to the usual matter term in Friedmann's equation, and identify the remainder as the DE energy density. Thus,

\ben\label{eq:rhoMNI1}
\frac{M_{P}^{-2}}{3H_{0}^{2}}\tilde{\rho}_{M} =(1-A)^{2}(1+z)^{3} \qquad
\text{and} \qquad
\label{eq:rhoXNI1}
\frac{M_{P}^{-2}}{3H_{0}^{2}}\tilde{\rho}_{X} = A^{2}+2\,A(1-A)(1+z)^{\frac{3}{2}}\, ,
\een
\no and from Eqs. (\ref{eq:rhoMX}.2), (\ref{eq:H1}) and (\ref{eq:rhoXNI1}) it follows that
\be\label{eq:c2(z)}
\tilde{c}^{2}=\frac{2A(1-A)(1+z)^{\frac{3}{2}}+A^{2}}{\left(A+(1-A)(1+z)^{\frac{3}{2}}\right)^{2}}\, .
\ee
In consistency with the findings of \cite{RadicellaPavon}, $\tilde{c}^{2}$ never decreases and tends from below to a constant value in the far future,
\be\label{eq:limC}
 \tilde{c}^{2}_{z\rightarrow\infty}= 0 \,,\qquad
 \tilde{c}^{2}_{z\rightarrow 0}= 1-(1-A)^{2} \,,\qquad
 \tilde{c}^{2}_{z\rightarrow -1}= 1\,,
\ee
\no where $0 \leq A \leq 1$ from observations. Table
\ref{tab:ex1a} shows the best fit values of the parameters and
their 1$\sigma$ errors. Table \ref{tab:ex1b} presents the
$\chi^{2}$ of model 1 and $\Lambda$CDM, obtained by fitting each
data set independently, and the total $\chi^{2}$. The obtained
$\chi^{2}$ per degree of freedom (dof) is $\chi^{2}/dof=1.00$.
Notice that $(1-A)^{2}=\tilde{\Omega}_{M 0}$ and its value
($\approx 0.17$) is  about $6\sigma$ lower than the one reported
by Komatsu \textit{et al.} \cite{Komatsu}. However, this value is
reached by using the last scattering sound horizon as a standard
ruler; i.e., it is not observed directly but by integrating the
background evolution. So truly, just a global background evolution
is obtained, which in the case of $\Lambda$CDM, gives the value of
$\Omega_{M 0}$ mentioned above, but in other models, as in this
one, it can vary. Figure \ref{fig:elipsesEx1} shows the 1$\sigma$
and 2$\sigma$ confidence regions and the best fit value of the
free parameters of the models.
\begin{table*}[!htb]
    \begin{tabular}{||p{4.2 cm}||p{2.5 cm} p{2.5 cm} p{2.5 cm} ||}
    \hline \hline
    $\,$ Model & $\;\;\;\;\;\;\;\;\;\Omega_{X\,0}$ &  $\;\;\;\;\;\;H_{0}$ & $\;\;\;\;\;\;\;\;A$ \\
    \hline \hline
    $\,$ Interacting holographic& $\; 0.73$  & $69.4\pm 1.7$ & $0.588 \pm 0.004$  \\
    \hline
    $\,$ $\tilde{c}^{2}$ holographic & $\; 0.830\pm 0.003$  & $69.4\pm 1.7$ & $0.588 \pm 0.004$  \\
    \hline
    $\,$ $\Lambda$CDM & $\; 0.720 \pm 0.003$  & $71.5^{+1.3}_{-1.5}$ & \qquad$\,$ ---  \\
    \hline \hline
    \end{tabular}
    \caption{\scriptsize Values of the parameters of the models obtained by constraining them
    with the observational data described in Sec. \ref{Constraints}. $\Omega_{X\,0}$
    is not a free parameter in any of the two holographic scenarios, but it is included
    for the sake of comparison with the conventional $\Lambda$CDM model. Notice that
    despite the fact that holographic models are described by the same Hubble function, they
    have different values of $\Omega_{X\,0}$. This point does not discard any model,
    since the value from WMAP7 \cite{Komatsu} is obtained using the last scattering
    surface as a standard ruler, i.e., integrating the $\Lambda$CDM background
    evolution from the last scattering surface to the present time.
    Thus, if the Hubble function is the same for different models, geometric
    data alone can not distinguish among them. The $H_{0}$ values are given
    in $km/s/Mpc$.}
    \label{tab:ex1a}
\end{table*}
\begin{table*}[!htb]
    \begin{tabular}{||p{2 cm}||p{1.2 cm} p{1.2 cm} p{1.2 cm} p{1.2 cm} p{1.2 cm} p{1.2 cm} p{1.5 cm}||}
    \hline \hline
    $\,$ Model & $\; \chi^{2}_{SN}$ &  $\chi^{2}_{BAO}$ & $\chi^{2}_{l_{A}}$ &   $\chi^{2}_{x-ray}$ &  $\chi^{2}_{H}$ & $\chi^{2}_{{\rm tot}}$ & $\chi^{2}_{{\rm tot}}/dof$ \\
    \hline \hline
    $\,$ Model 1 & $\; 554.8$  & $1.7$ & $0.3$ & $44.9$ & $11.3$ & $613.0$ & $\; \; 1.00$ \\
    \hline
    $\,$ $\Lambda$CDM & $\; 542.7$  & $1.2$ & $0.7$ & $42.3$ & $8.8$ & $595.7$ & $\; \; 0.97$ \\
    \hline \hline
    \end{tabular}
    \caption{\scriptsize $\chi^{2}$ values for the holographic model of Sec. \ref{sec:EX1} and the $\Lambda$CDM model. Each of them has two free parameters ($A$ and $H_{0}$ the holographic, and $\Omega_{X\, 0}$ and  $H_{0}$ the $\Lambda$CDM).}
    \label{tab:ex1b}
\end{table*}
\\Although H(z) coincides with the corresponding expression in \cite{PavonDuranZimdahl}, $\rho_{M}$, $\rho_{X}$, and $w(z)$ do not. The functional form of $\tilde{w}$ coincides with that of the EoS parameter for the interacting case - cf. Equation (\ref{eq:wInt})-,
\be\label{eq:wNoInt}
\tilde{w}=-\frac{A}{\tilde{\Omega}_{X}}\frac{H_{0}}{H}
\ee
but has a different time dependence since $\Omega_{X}\neq\tilde{\Omega}_{X}$, as the left panel of Fig. \ref{fig:ex1} shows. Notice that in the interacting case, $w$ crosses the phantom divide line ($w=-1$). However, the $w_{eff}$ defined in the line below Eq. (\ref{eq:evolEqNoIntMX}), does not cross the said line. The pressure, obtained by multiplying Eq. (\ref{eq:wNoInt}) by Eq. (\ref{eq:rhoMX}.2) and using Eq. (\ref{eq:H1}), described here for later purposes, reads
\be\label{eq:Pressure}
\tilde{P}=\tilde{w}\tilde{\rho}_{X}=-3AM_{P}^{2}H_{0}^{2}\left[A+(1-A)(1+z)^{\frac{3}{2}}\right] \,.
\ee
As shown in the right panel of Fig. \ref{fig:ex1}, the coincidence problem is solved (i.e., $r$ is constant) in the interacting case (solid green line). By contrast, in the $\tilde{c}^{2}$ model (thin dot-dashed red lines), it is not solved but is much less severe than in the $\Lambda$CDM model (thick short dashed blue line).
\section {Model 2}\label{sec:EX2}
\begin{figure*}[!htb]
  \begin{center}
    \begin{tabular}{cc}
      \resizebox{80mm}{!}{\includegraphics{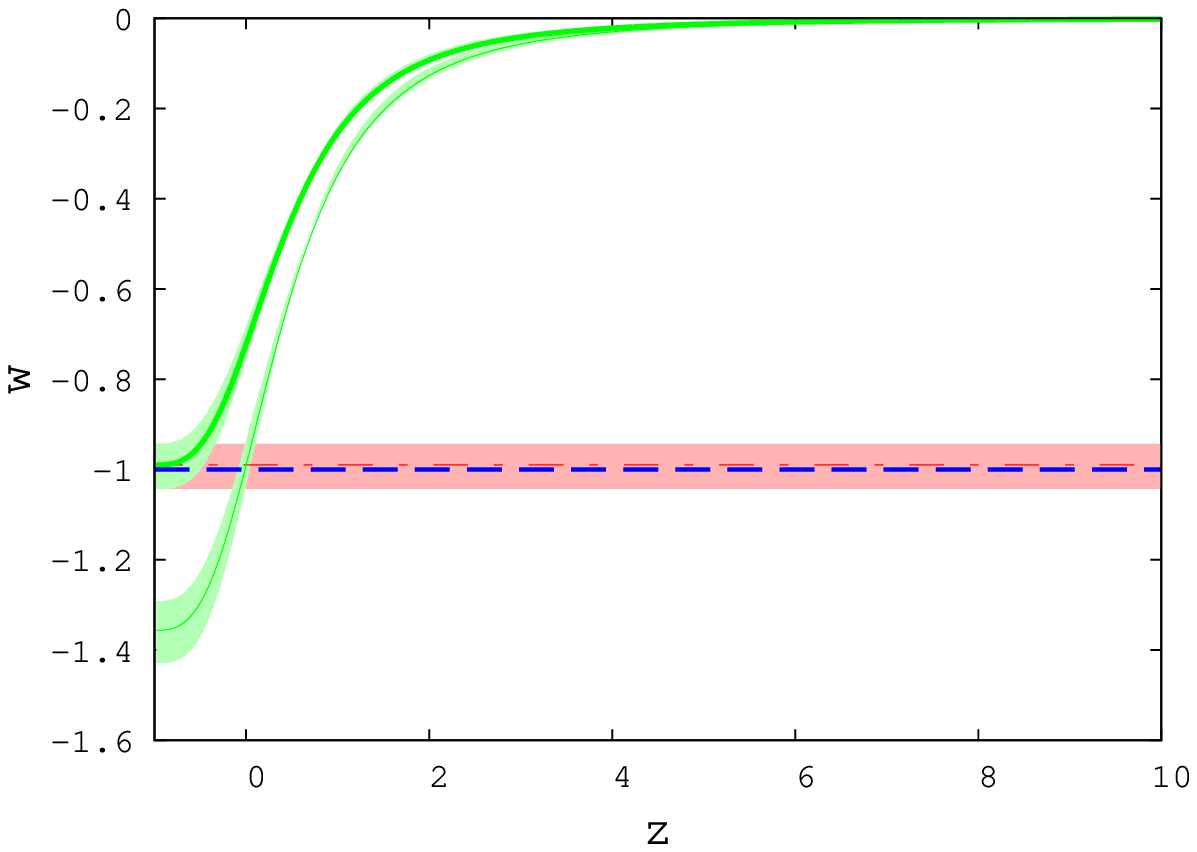}}&
      \resizebox{80mm}{!}{\includegraphics{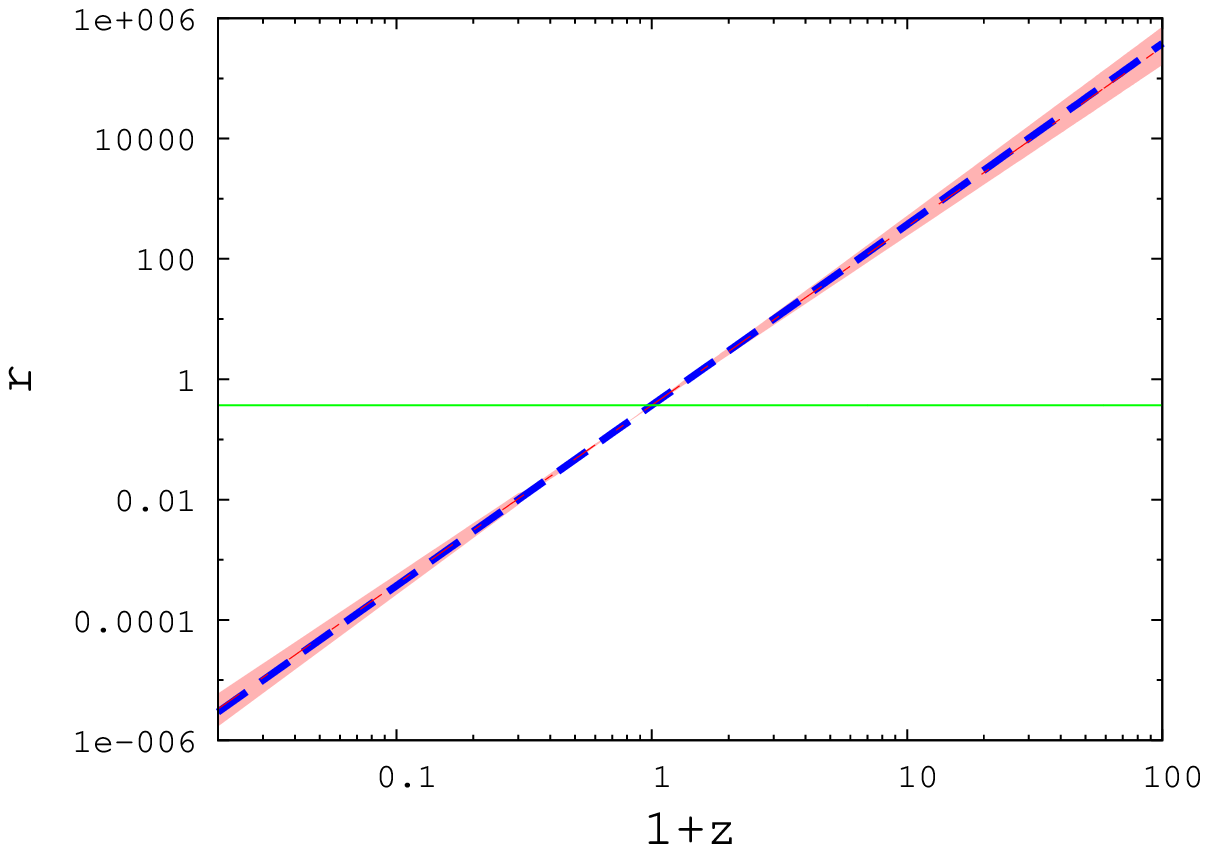}}\\
    \end{tabular}
    \caption{\scriptsize Left panel: EoS parameter for the interacting ($w$ with thin line, and $w_{eff}$ with thick line), the $\tilde{c}^{2}$  and $\Lambda$CDM models. Right panel: energy densities ratios, $r\equiv\rho_{M}/\rho_{X}$, versus $1+z$ for the $\Lambda$CDM , the interacting and the $\tilde{c}^{2}$ models. Notice that the energy densities ratio of the $\tilde{c}^{2}$ and $\Lambda$CDM models practically overlap. All graphs were drawn using the best fit values of the respective parameters, shown in Table \ref{tab:ex2a}. Solid (green) lines are used for the interacting case, thin dot-dashed (red) lines for the $\tilde{c}^{2}$, and thick short dashed blue for the $\Lambda$CDM. The 1$\sigma$ region of the parameters is also plotted.}
    \label{fig:ex2}
 \end{center}
\end{figure*}

\begin{figure*}[!htb]
  \begin{center}
    \begin{tabular}{cc}
      \resizebox{80mm}{!}{\includegraphics{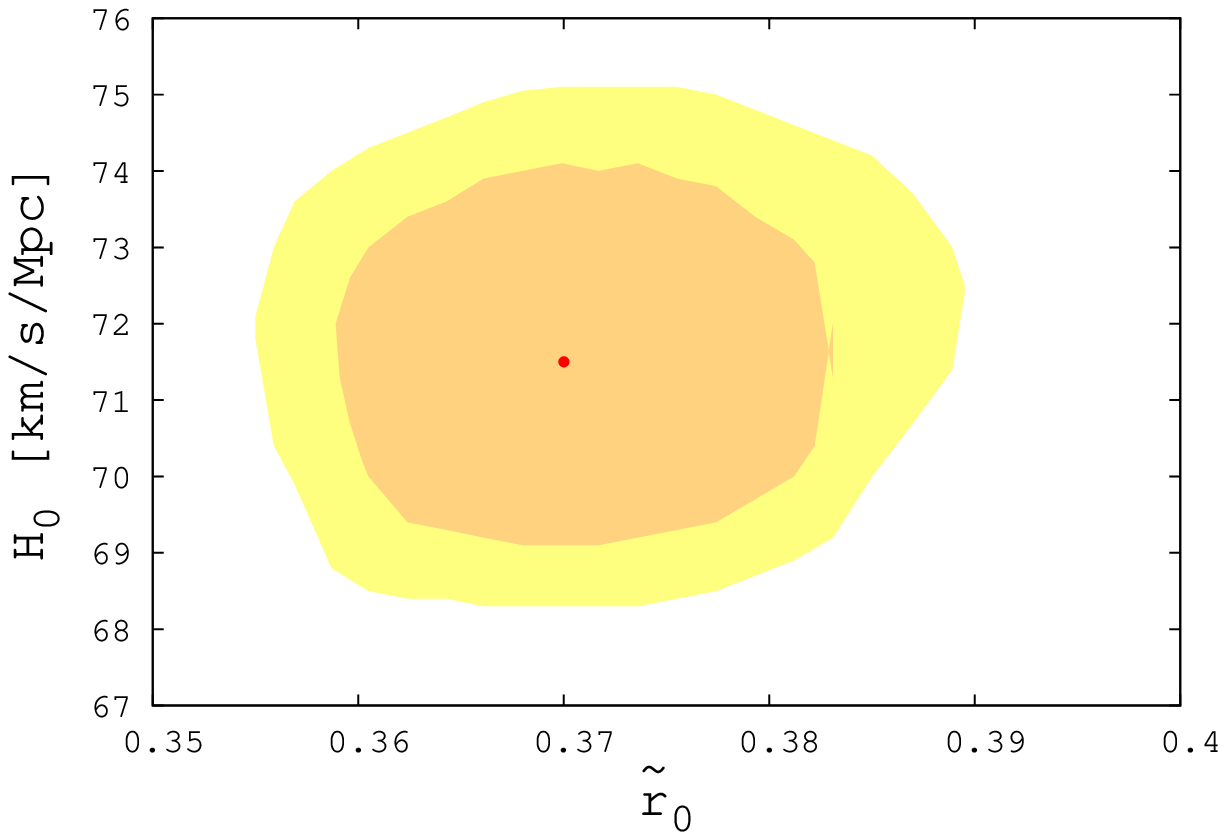}}&
      \resizebox{80mm}{!}{\includegraphics{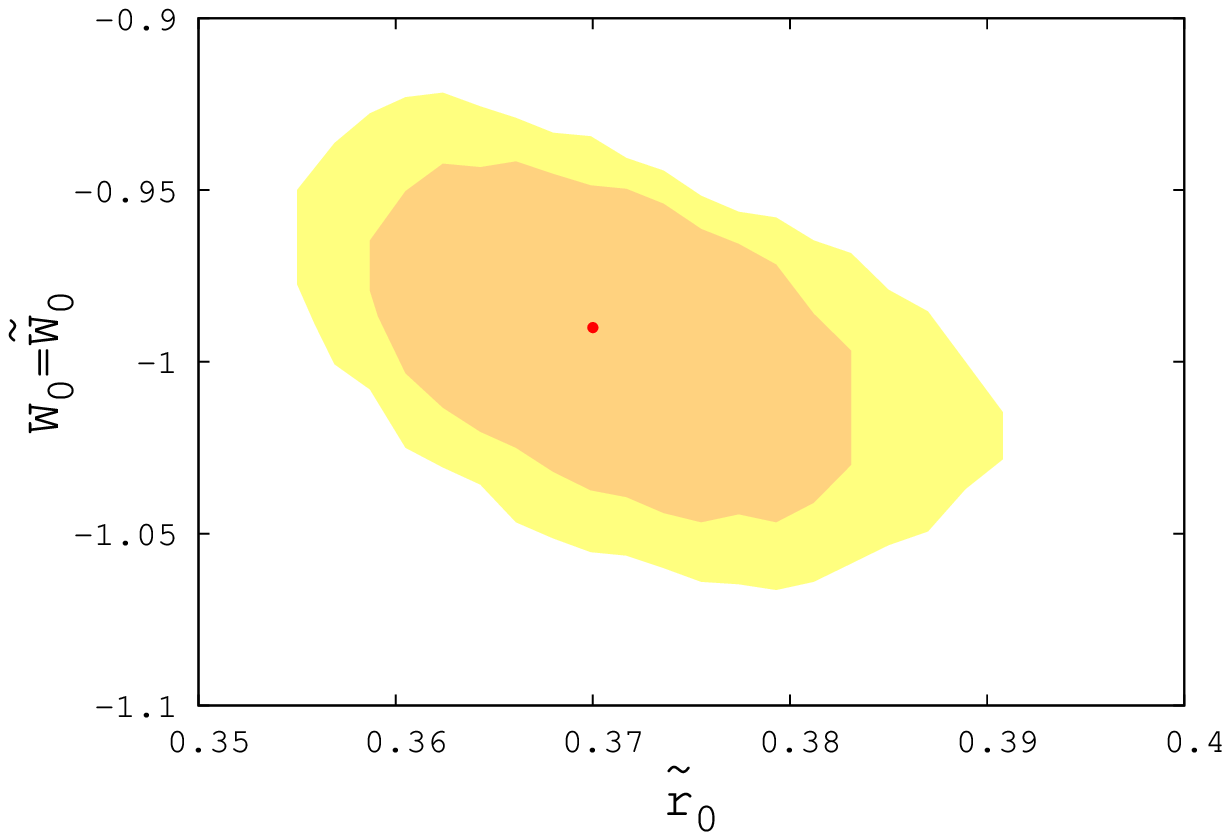}}\\
    \end{tabular}
    \caption{\scriptsize Left panel: The 1$\sigma$ and 2$\sigma$ confidence regions for the parameters $\tilde{r}_{0}$ and $H_{0}$ of model 2. Right panel: The same regions for the parameters $w_{0}$ (equivalent to $\tilde{w}_{0}$ and derived from $\epsilon=-3w_{0}$) and $\tilde{r}_{0}$. Notice that $\tilde{r}_{0}$ is the ratio of the energy densities just for the $\tilde{c}^{2}(t)$ (noninteracting) case; for the interacting case it has no straightforward physical meaning. Central dots indicate the best fit values.}
    \label{fig:elipsesEx2}
 \end{center}
\end{figure*}

In this model, DM and DE evolve separately (i.e., $Q=0$) but the holographic parameter $\tilde{c}^{2}$ varies slowly with time. To have $0 \leq \tilde{c}^{2} \leq 1$, and $\left(\tilde{c}^{2}\right)\dot{}\geq 0$, we use the parametrization
\be\label{eq:cDeT4}
\tilde{c}^{2}=\frac{1}{1+\tilde{r}_{0}(1+z)^{\epsilon}}
\ee
\no where $\tilde{r}_{0}\equiv\frac{\tilde{\Omega}_{M0}}{\tilde{\Omega}_{X0}}$ and $\epsilon$ is a semipositive definite constant (notice that $\tilde{c}^{2}(z=0)=\tilde{\Omega}_{X0}$). In this case, the Hubble function
\be\label{eq:H4}
H=H_{0}\sqrt{\tilde{\Omega}_{M0}(1+z)^{3}+\tilde{\Omega}_{X0}(1+z)^{3-\epsilon}}
\ee
\no coincides with the Hubble function of the spatially flat $w$CDM model with $\tilde{w}=-\frac{\epsilon}{3}$. It obviously reduces to the $\Lambda$CDM model for $\epsilon=3$.
If we consider the Hubble function in  Eq.(\ref{eq:H4}) as coming from an interaction between DE and DM, by Eq. (\ref{eq:wi}) the interacting term would be
\be\label{eq:Q4}
Q=-3\,c^{2}\,w\rho_{M}H \,\text{,}
\ee
\no where the EoS parameter of the interacting case is $w=-\frac{\epsilon}{3c^{2}\left(1+\tilde{r}_{0}(1+z)^{\epsilon}\right)}$.\\
The variation of $\tilde{c}^{2}$ breaks the holographic dependence of DE density. But as for the rate of variation of $\tilde{c}^{2}$ we must have $\left(\tilde{c}^{2}\right)\dot{}/c^{2}=\epsilon\tilde{r}_{0}(1+z)^{\epsilon}H/(1+\tilde{r}_{0}(1+z)^{\epsilon})\lesssim H$, it may be considered as a small variation in the level of saturation. For the last inequality to be fulfilled, $f(z)\equiv\,\epsilon\Omega_{M0}(1+z)^{\epsilon}/(\Omega_{X0}+\Omega_{M0}(1+z)^{\epsilon})\leq 1$. This was not always true in the past, but the maximum value of $f(z)$ , that is monotonously decreasing, is $f(z\rightarrow\infty)=\epsilon$. Using the best fit value for $\epsilon$, shown in Table \ref{tab:ex2a}, its maximum variation rate is of the order of the expansion rate, and the model can still be considered holographic. The $\chi^{2}$ values obtained by fitting the model with the different data sets, are shown in Table \ref{tab:ex2a}. The $\chi^{2}$ per dof obtained with the fit of all the data sets together is $\chi^{2}/dof=0.97$.
\\
The left panel of Fig. \ref{fig:ex2} shows that the EoS parameter $w$ (thin solid green line) crosses the phantom divide line but the effective one $w_{eff}$ (thick solid green line) of the interacting case does not. The noninteracting case, as we have mentioned before, is just a $w$CDM model with $w=-\frac{\epsilon}{3}$ (thin dot-dashed red line). The right panel of Fig. \ref{fig:ex2} shows that the interacting model (solid green line) solves the coincidence problem, and in the case of the $\tilde{c}^{2}$ one (thin dot-dashed red line), it overlaps the $\Lambda$CDM line (thick short dashed blue line), since $w\gtrsim-1$. Figure \ref{fig:elipsesEx2} depicts the 1$\sigma$ and 2$\sigma$ regions for the parameters $\tilde{r}_{0}$ and $H_{0}$ (left panel) and $\tilde{r}_{0}$ and $w_{0}$ (right panel). Both panels are the same for both scenarios, the interacting and the $\tilde{c}^{2}$, however, $\tilde{r}_{0}$ is only related to the DE and DM densities in the $\tilde{c}^{2}$ description; in the interacting case it has no physical meaning. In both cases today's value of the EoS is the same $w_{0}=\tilde{w}_{0}=-\frac{\epsilon}{3}$, as the left panel of Fig. \ref{fig:ex2} shows.
\begin{table*}[!htb]
    \begin{tabular}{||p{4.2 cm}||p{2.7 cm} p{2.3 cm} p{2.5 cm} p{2.0 cm} ||}
    \hline \hline
    $\,$ Model & $\;\;\;\;\;\;\;\;\Omega_{X\,0}$ &  $\;\;\;\;\;\;H_{0}$ & $\;\;\;\;\;\;\;\;\tilde{r}_{0}$ & $\;\;\;\;\;\;\;\epsilon$ \\
    \hline \hline
    $\,$ Interacting holographic& $\; 0.73$  & $71.5 \pm 2.6$ & $0.370\pm 0.013$ & $2.97^{+ 0.16}_{- 0.14}$  \\
    \hline
    $\,$ $\tilde{c}^{2}$ holographic & $\; 0.730 \pm 0.007$ & $71.5 \pm 2.6$ & $0.370\pm 0.013$ & $2.97^{+0.16}_{-0.14}$  \\
    \hline
    $\,$ $\Lambda$CDM & $\; 0.720 \pm 0.003$  & $71.5^{+1.3}_{-1.5}$ & \qquad$\,$ --- & \quad$\;$ --- \\
    \hline \hline
    \end{tabular}
    \caption{\scriptsize Best fit values of the free parameters of the models. In the  $\tilde{c}^{2}$ holographic scenario, $\Omega_{X\,0}$ is obtained from the free parameter $\tilde{r}_{0}$, so the model has only three free parameters. In the interacting one, for $\Omega_{X\,0}$ we use the value obtained in \cite{Komatsu}. The $H_{0}$ values are given in $km/s/Mpc$.}
    \label{tab:ex2a}
\end{table*}
\begin{table*}[!htb]
    \begin{tabular}{||p{2 cm}||p{1.2 cm} p{1.2 cm} p{1.2 cm} p{1.2 cm} p{1.2 cm} p{1.2 cm} p{1.5 cm}||}
    \hline \hline
    $\,$ Model & $\; \chi^{2}_{SN}$ &  $\chi^{2}_{BAO}$ & $\chi^{2}_{l_{A}}$ &   $\chi^{2}_{x-ray}$ &  $\chi^{2}_{H}$ & $\chi^{2}_{{\rm tot}}$ & $\chi^{2}_{{\rm tot}}/dof$ \\
    \hline \hline
    $\,$ Model 2 & $\; 542.7$  & $1.2$ & $0.0$ & $41.5$ & $8.8$ & $594.2$ & $\; \; 0.97$ \\
    \hline
    $\,$ $\Lambda$CDM & $\; 542.7$  & $1.2$ & $0.7$ & $42.3$ & $8.8$ & $595.7$ & $\; \; 0.97$ \\
    \hline \hline
    \end{tabular}
    \caption{\scriptsize $\chi^{2}$ values for the holographic model studied in Sec. \ref{sec:EX2} and for $\Lambda$CDM model. In the former, the free parameters are $\tilde{r}_{0}$, $\epsilon$ and $H_{0}$. In the latter the free parameters are two, $\Omega_{X\, 0}$ and  $H_{0}$. }
    \label{tab:ex2b}
\end{table*}
\section {Evolution of the subhorizon perturbations}\label{Perturbations}
We have seen that in model 1 (Sec. \ref{sec:EX1}), the interacting version is preferred over the noninteracting one because $\tilde{\Omega}_{M 0} \approx 0.17$, too small as compared with results of Komatsu et \textit{al}. \cite{Komatsu}. In model 2 (Sec. \ref{sec:EX2}), both seem to be compatible with observations at the background level. A further study should be done, in general, to establish which scenario is observationally favored. Here, we make a simple perturbation study, just considering matter perturbations inside the horizon. This study just intends to illustrate that a perturbative analysis can permit us to distinguish the interacting from the noninteracting scenario, despite the fact that they share the same $H(z)$.

Once the Universe becomes matter dominated, the anisotropic stress due to neutrinos will be negligible and, in the Newtonian gauge, it will suffice just one single gravitational potential -say $\phi$- to determine the flat metric element
\be\label{eq:PertrubedMetric}
ds^{2} = -(1 + 2\phi)dt^{2} + a^{2}(1 - 2\phi)dx^{i}dx_{i} .
\ee
In the interacting case, the energy-momentum tensors of DM and DE are not independently conserved. For the matter component, by perturbing the conservation equation $T^{\mu \,\nu}_{M\;\; ;\mu}=Q^{\mu}_{M}$, where $Q^{\mu}_{M}\equiv Q u_{M}^{\mu}$ (with $u_{M}^{\mu}$ the four velocity of the matter component, and no momentum transfer in the DM reference frame is assumed) \cite{KodamaSasaki,Maartens}, the equations of motion for the energy density and the velocity divergence ($\theta=i k^{j}v_{j}$) perturbations, with the speed of sound, $c_{s\,M}^{2}=0$, and the EoS parameter, $w_{M}=0$, are
\ben\label{eq:PertrubedDensity}
    \dot{\delta}_{M}&=&-\frac{\theta_{M}}{a}+ 3\dot{\phi}+\frac{\delta Q}{\rho_{M}}+ \frac{Q}{\rho_{M}}\left(\phi-\delta_{M}\right) \;\text{,}\\
    \label{eq:PertrubedVelocity}
    \dot{\theta}_{M}&=&-H\theta_{M}+\frac{k^{2}}{a}\phi \;\text{,}
\een
\no and for the DE component $T^{\mu \,\nu}_{X\;\; ;\mu}=Q^{\mu}_{X}$, where $Q^{\mu}_{X}\equiv - Q u_{M}^{\mu}$. Since DE can be considered a scalar field, the effective speed of sound $c_{s\,X}^{2}=1$, the dynamical equations for perturbations are
\begin{widetext}
\ben\label{eq:PertrubedDEDensity}    \dot{\delta}_{X}&=&-\left(1+w\right)\frac{\theta_{X}}{a}-3H\left(1-w\right)\delta_{X}-3aH\left[3H\left(1-w^{2}\right)+\dot{w}_{X}\right]\frac{\theta_{X}}{k^{2}}\\ \nonumber &+&3\left(1+w\right)\dot{\phi}-\frac{Q}{\rho_{X}}\left[\phi-\delta_{X}+3aH\left(1-w\right)\frac{\theta_{X}}{k^{2}}\right]-\frac{\delta Q}{\rho_{X}} \;\text{,}\\
    \label{eq:PertrubedDEVelocity}    \dot{\theta}_{X}&=&2H\theta_{X}+\frac{1}{\left(1+w\right)}\frac{k^{2}}{a}\delta_{X}-\frac{k^{2}}{a}\phi-\frac{Q}{\left(1+w\right)\rho_{X}}\left(\theta_{M}-2\theta_{X}\right)\;\text{,}
\een
\end{widetext}
\no while the Fourier transformed time-time and space-time components of the Einstein equations are
\ben\label{eq:Einstein00}
    \frac{k^{2}}{a^{2}}\phi+3H(\dot{\phi}+H\phi)&=&-4\pi G \sum_{i} \rho_{i}\,\delta_{i} \;\text{,}\\
\label{eq:Einstein0i}
   \frac{k^{2}}{a}\left(\dot{\phi}+H\phi\right)&=&4\pi G\sum_{i} (\rho_{i}+P_{i})\,\theta_{i} \;\text{.}
\een
A more detailed derivation of the perturbation equations can be found in \cite{Maartens}. As we are concerned with subhorizon scales, we just consider the case in which $k \gg aH$, and so, the Newtonian limit of Poisson's equation (\ref{eq:Einstein00}) is just
\be\label{eq:Poisson}
    \frac{k^{2}}{a^{2}}\phi=-4\pi G \left(\rho_{M}\delta_{M}+\rho_{X}\delta_{X}\right) \;\text{.}
\ee
From  Eq.(\ref{eq:Poisson}) and bearing in mind that every single energy component obeys $\frac{8\pi G}{3}\rho_{i}\leq H^{2}$, one sees that the gravitational potential (and its derivatives) can be neglected when compared with the density perturbations. After all this, Eqs. (\ref{eq:PertrubedDensity}), (\ref{eq:PertrubedVelocity}), (\ref{eq:PertrubedDEDensity}) and (\ref{eq:PertrubedDEVelocity}) simplify to
\ben\label{eq:PertrubedDensityNew}
\dot{\delta}_{M}&=&-\frac{\theta_{M}}{a}\;\text{,}\\
\label{eq:PertrubedVelocityNew}
\dot{\theta}_{M}&=&-H\theta_{M}+\frac{k^{2}}{a}\phi \;\text{,}
\een
\ben
\label{eq:PertrubedDEDensityNew}
\dot{\delta}_{X}&=&-\left(1+w\right)\frac{\theta_{X}}{a}-3H\left(1-w\right)\delta_{X}+\frac{1}{\rho_{X}}\left(Q\delta_{X}-\delta Q\right) \;\text{,}\\
\label{eq:PertrubedDEVelocityNew}    \dot{\theta}_{X}&=&\frac{1}{\left(1+w\right)}\frac{k^{2}}{a}\delta_{X}-\frac{Q}{\left(1+w\right)\rho_{X}}\left(\theta_{M}-2\theta_{X}\right)\;\text{,}
\een
Notice that in the interacting version of model 1, $\delta Q=3AH_{0}\rho_{M}\delta_{M}$, but in model 2, since $Q$ includes a dependence in $w$ and $H$, things are more involved. However, we assume, as in \cite{Maartens}, that the product $3c^{2}wH$ is just an approximation to a time (but not position) dependent interaction rate, so there are no perturbations on it, and $\delta Q=-3c^{2}wH\rho_{M}\delta_{M}$. In the $\tilde{c}^{2}$ scenarios of both models $Q=0$ and $\delta Q=0$.
\subsection {Initial conditions}\label{InitialConditions}
To solve numerically these four coupled differential equations, we must choose some initial conditions for the density and velocity perturbations. We set them at $z_{i}=1000$. We impose the potential to be a constant (we have seen that it and its time derivative are much smaller than density perturbations), and using the perturbed Einstein equations, (\ref{eq:Einstein0i}) and (\ref{eq:Poisson}), we find the density and velocity initial conditions
\ben\label{eq:delta_i}
    k^{2}\phi&=& -\frac{3}{2(1+z_{i})^{2}}\left(\rho_{M\,i}\,\delta_{M}(z_{i},k)+\rho_{X\,i}\,\delta_{X}(z_{i},k)\right) \;\text{,}\\
\label{eq:theta_i}
    k^{2}\phi&=&\frac{3}{2(1+z_{i})H}\left(\rho_{M\,i}\,\theta_{M}(z_{i},k)+\rho_{X\,i}\,\theta_{X}(z_{i},k)\right)\;\text{.}
\een
\no for each model. Since we obtain, numerically, that the
evolution of perturbations is nearly independent of the wavenumber
$k$ in the range $0.001\; h\, Mpc^{-1}\leq k \leq 0.1\; h\,
Mpc^{-1}$, which includes all the interesting scales under
consideration, we shall assume that initially DE perturbations are
proportional to the DM perturbation, i.e.,
$\delta_{X}(z_{i},k)=\alpha\,\delta_{M}(z_{i},k)$ and
$\theta_{X}(z_{i},k)=\beta\,\theta_{M}(z_{i},k)$, with $\alpha $ a
nonnegative constant.
\subsubsection {Interacting version of model 1}\label{ModelIint}
We have freedom to normalize the matter density contrast as $\delta_{M}(z_{i})=\frac{1}{1+z_{i}}$, and 
then by Eq. (\ref{eq:Einstein0i}) and (\ref{eq:Poisson}), we find $\theta_{M}(z_{i})=-\left(\frac{1+\alpha\,\Omega_{X 0}}{\Omega_{M 0}+\beta\,\Omega_{X 0}}\right)\frac{H}{(1+z_{i})^{2}}$. For the DE component we take different options for both $\alpha$ and $\beta$.  The dashed (green) lines of the left panel of Fig. \ref{fig:deltaGrowth} correspond to the noninteracting scenarios. The solid (blue) line is for the $\Lambda$CDM model, shown for comparison purposes. Each different dashed (green) line depicts various initial conditions, from top to bottom, $\alpha=0$ and $\beta=0$, $\alpha=1$ and $\beta=1$, and $\alpha=1$ and $\beta=10$. Notice that the interacting scenarios exhibit a dependence on the initial conditions chosen for the DE component. The evolution of perturbations depend on the initial conditions on $\theta_{X}$ but not so much on $\delta_{X}$.
\subsubsection {$\tilde{c}^{2}$ version of model 1}\label{ModelIct}
For the matter component we impose as in the previous case $\delta_{M}(z_{i})=\frac{1}{1+z_{i}}$ and find $\theta_{M}(z_{i})=-\left(\frac{1+2\alpha\,A\,(1-A)\,(1+z)^{-\frac{3}{2}}}{1+2\beta\,A\,(1-A)\,(1+z)^{-\frac{3}{2}}}\right)\frac{H}{(1+z_{i})^{2}}$. The  left panel of Fig. \ref{fig:deltaGrowth} shows the evolution of $\delta_{M}$, dot-dashed (red) lines, for different options for both $\delta_{X}(z_{i})$ and $\theta_{X}(z_{i})$. Each different dot-dashed (red) line depicts two initial conditions, $\alpha=0$ and $\beta=0$ and $\alpha=1$ and $\beta=10$ (they practically overlap). In this scenario, the final result does not depend on the chosen initial conditions.
\subsubsection {Interacting version of model 2}\label{ModelIIint}
For the matter component we obtain as in the previous section \ref{ModelIint} $\delta_{M}(z_{i})=\frac{1}{1+z_{i}}$ and  $\theta_{M}(z_{i})=-\left(\frac{1+\alpha\,\Omega_{X 0}}{\Omega_{M 0}+\beta\,\Omega_{X 0}}\right)\frac{H}{(1+z_{i})^{2}}$. For the DE component we take different options for both $\delta_{X}(z_{i})$ and $\theta_{X}(z_{i})$.  The dashed (green) lines of the right panel of Fig. \ref{fig:deltaGrowth} correspond to the noninteracting scenario. The solid (blue) line is for the $\Lambda$CDM model, shown for comparison purposes. Each different dashed (green) line depicts, from top to bottom in the right, $\alpha=0$ and $\beta=0$, $\alpha=1$ and $\beta=1$ and $\alpha=1$ and $\beta=10$. Notice that the interacting scenarios exhibit a dependence on the initial conditions chosen for the DE component.
\subsubsection {$\tilde{c}^{2}$ version of model 2}\label{ModelIIct}
For the matter component we find $\delta_{M}(z_{i})=\frac{1}{1+z_{i}}$ and  $\theta_{M}(z_{i})=-\frac{H}{(1+z_{i})^{2}}$. The right panel of Fig. \ref{fig:deltaGrowth}, shows the evolution of $\delta_{M}$, with dot-dashed (red) lines, for various options for both $\delta_{X}(z_{i})$ and $\theta_{X}(z_{i})$. Each different dot-dashed (red) line depicts the initial conditions, $\alpha=0$ and $\beta=0$, and $\alpha=1$ and $\beta=10$, though they practically overlap. Again, the final result does not depend on the initial conditions chosen.

\subsection {Results}\label{Results}
\begin{figure*}[!htb]
  \begin{center}
    \begin{tabular}{cc}
      \resizebox{80mm}{!}{\includegraphics{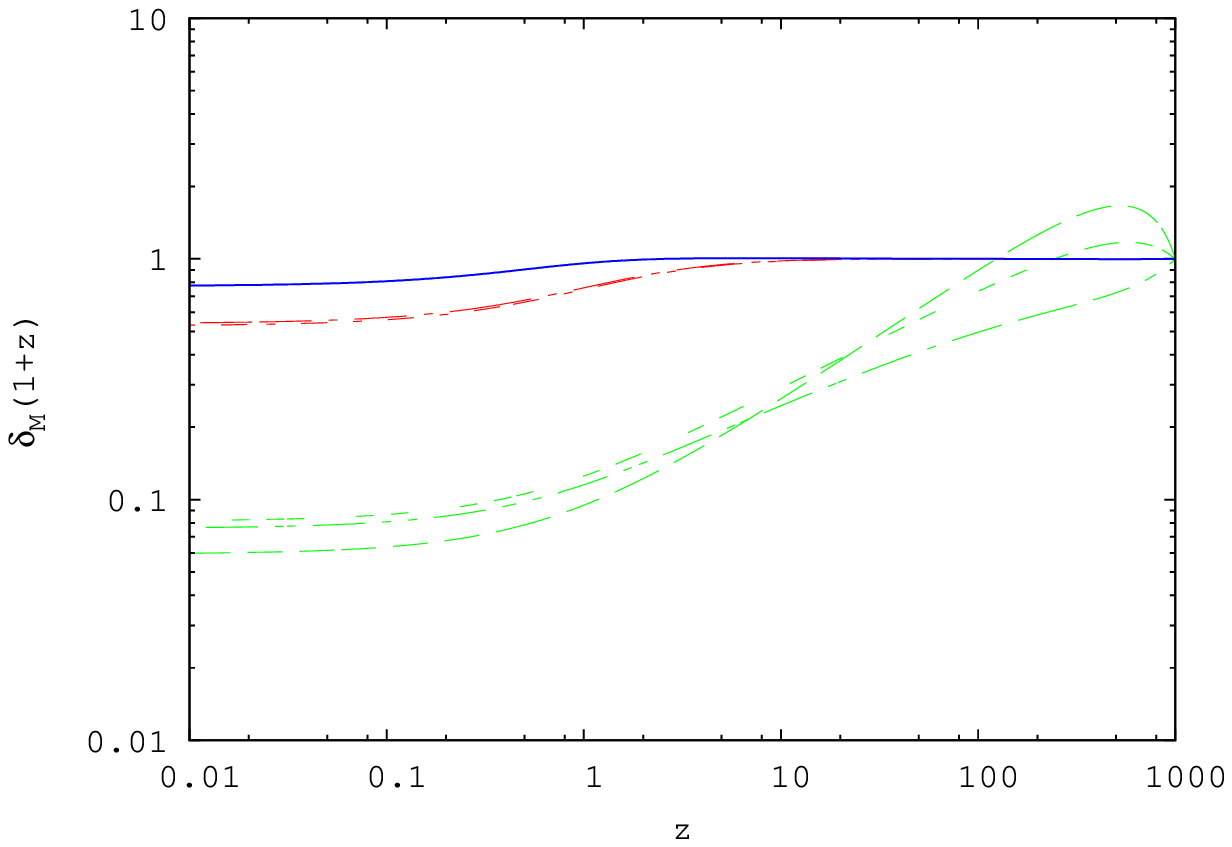}}&
      \resizebox{80mm}{!}{\includegraphics{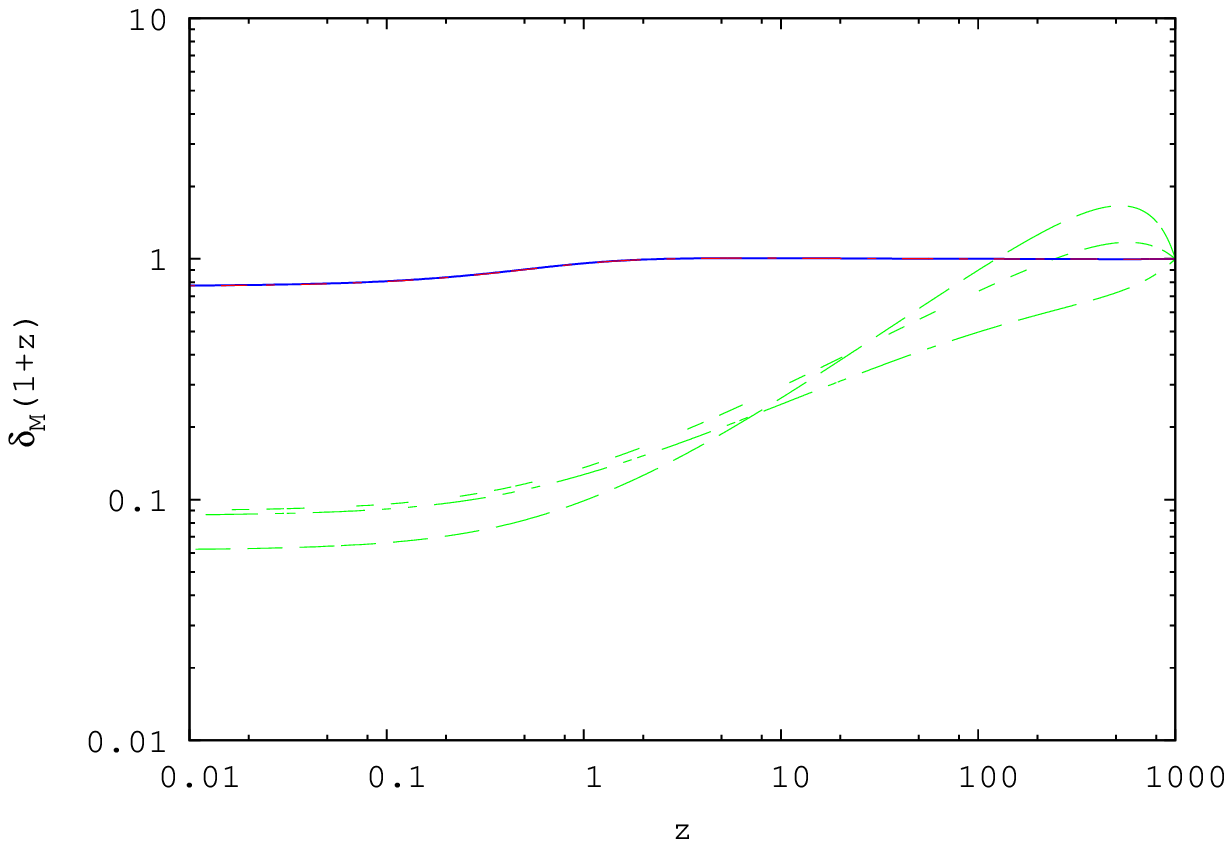}}\\
    \end{tabular}
    \caption{\scriptsize Left panel: The evolution of DM density perturbations versus redshift in model 1. Right panel: The same for model 2. In plotting the graphs the dashed (green) lines describe the matter density perturbations of the interacting scenario, the dot-dashed (red) lines the $\tilde{c}^{2}$, and the solid (blue) line $\Lambda$CDM for comparison purposes. Different initial conditions are used for the interacting versions; from top to bottom at the right, $\alpha=0$ and $\beta=0$, $\alpha=1$ and $\beta=1$ and $\alpha=1$ and $\beta=10$. For the $\tilde{c}^{2}$, two different sets of initial conditions are used, but the corresponding graphs practically overlap each other ($\alpha=0$ and $\beta=0$, and $\alpha=1$ and $\beta=10$). Notice that in the right panel, the $\tilde{c}^{2}$ scenario overlaps $\Lambda$CDM, as it behaves as a $w$CDM model with $w=0.99$.}
    \label{fig:deltaGrowth}
 \end{center}
\end{figure*}

\begin{figure*}[!htb]
  \begin{center}
    \begin{tabular}{cc}
      \resizebox{80mm}{!}{\includegraphics{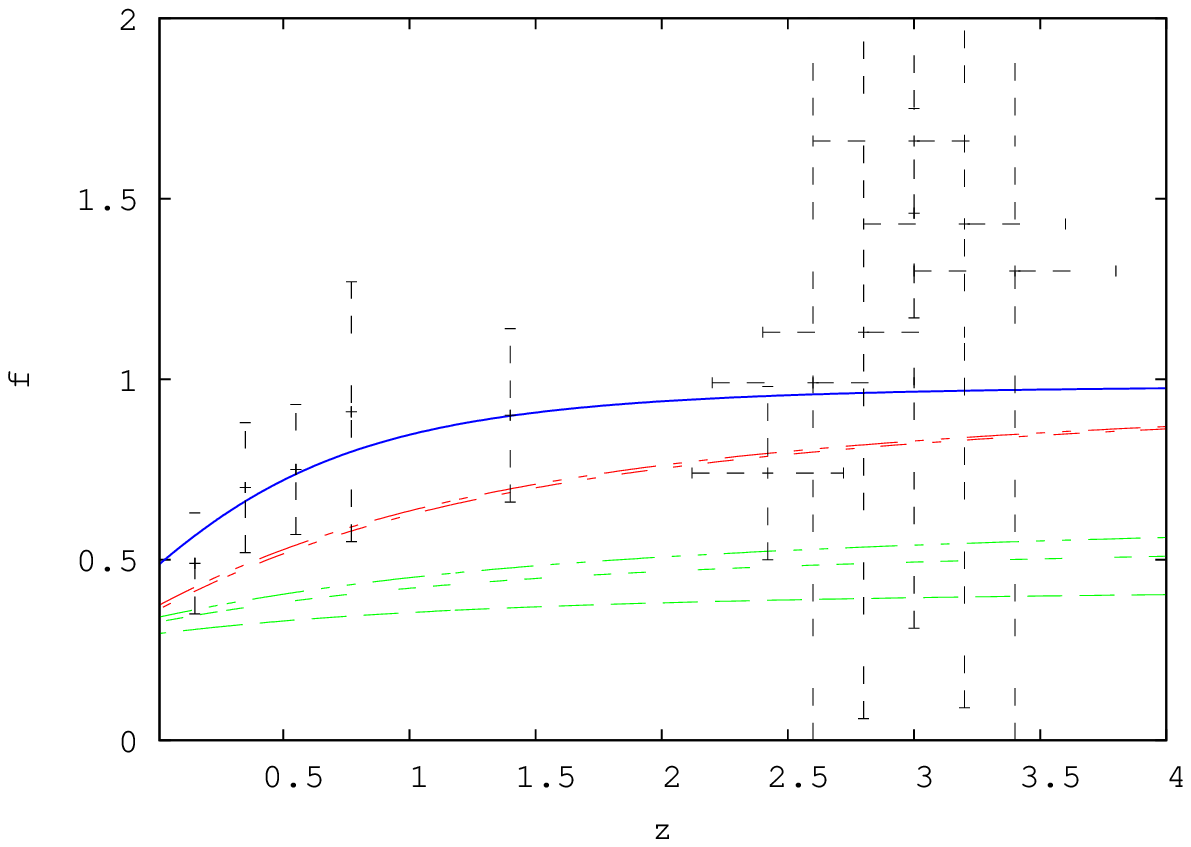}}&
      \resizebox{80mm}{!}{\includegraphics{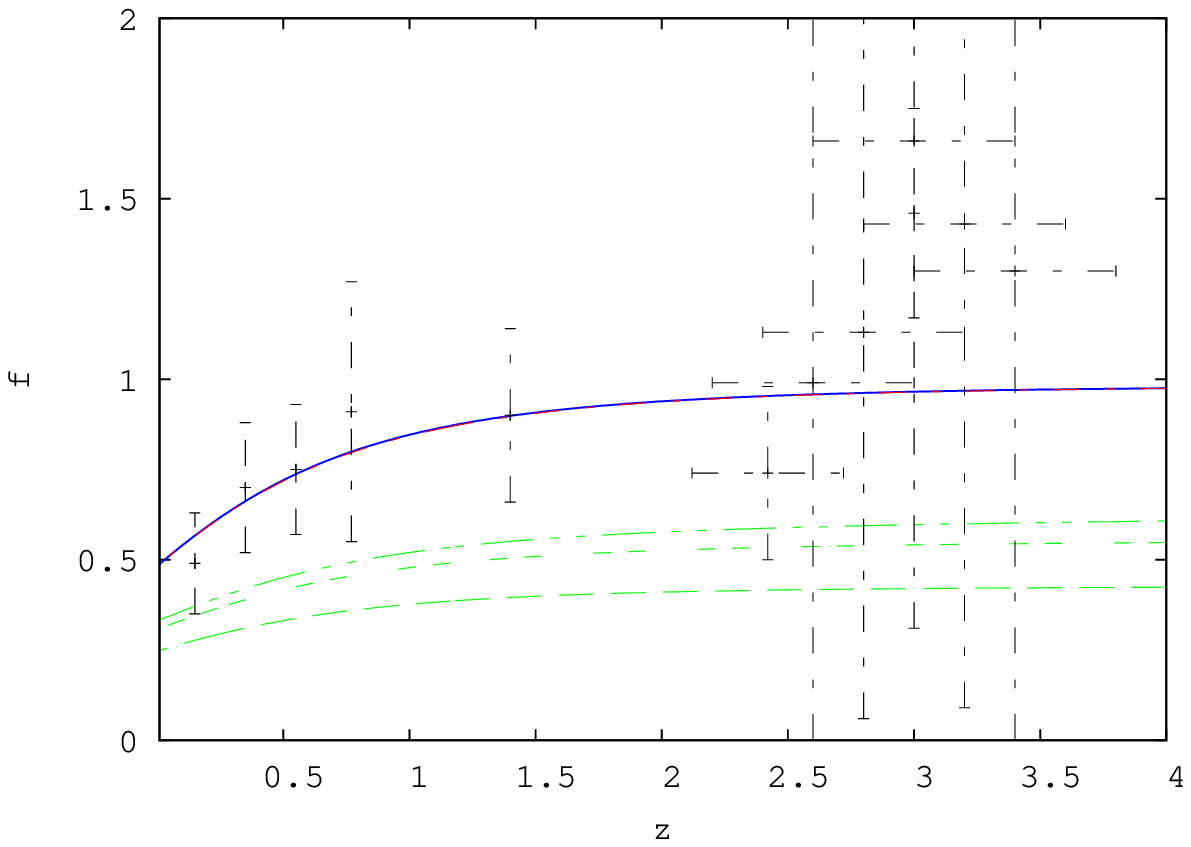}}\\
    \end{tabular}
    \caption{\scriptsize Left panel: the evolution of the growth function, $f\equiv \frac{d\ln \delta_{M}}{d \ln a}$, versus redshift for model 1. Right panel: the same for model 2. In plotting the graphs the dashed (green) lines describe the growth function of the interacting scenario, the dot-dashed (red) lines the $\tilde{c}^{2}$, and the solid (blue) line $\Lambda$CDM for comparison purposes. Different initial conditions are used for the interacting versions; from top to bottom $\alpha=0$ and $\beta=0$, $\alpha=1$ and $\beta=1$ and $\alpha=1$ and $\beta=10$. For the $\tilde{c}^{2}$, two different sets of initial conditions are used, but they practically overlap each other ($\alpha=0$ and $\beta=0$, and $\alpha=1$ and $\beta=10$). Notice that in the right panel, the $\tilde{c}^{2}$ scenario overlaps $\Lambda$CDM, as it behaves as a $w$CDM model with $w=0.99$. Observational data is borrowed from \cite{Gong}.}
    \label{fig:f}
 \end{center}
\end{figure*}

Figure \ref{fig:deltaGrowth} shows the numerical solution for $\delta_{M}$ for model 1 (left panel), and  for model 2 (right panel), in both cases for $k=0.01\; h\, Mpc^{-1}$. However, the outcome is quite independent of the wavenumber $k$ in the range $0.001\; h\, Mpc^{-1}\leq k \leq 0.1\; h\, Mpc^{-1}$ that includes all the interesting scales under consideration. In the interacting case, the matter density perturbations do not depend much on initial conditions imposed on the $\delta_{X}$, but they do on the initial conditions on $\theta_{X}$. Notice that in any case, the matter density perturbations clearly differ in both scenarios, the interacting and the $\tilde{c}^{2}$ one. The most favored (the closer to $\Lambda$CDM), at least in the two models studied here, are the $\tilde{c}^{2}$ scenarios, since low density perturbations at $z\approx 10$ can be problematic for the large scale structure formation.
To confront it with observations, we resort to the growth function, $f\equiv d \ln \delta_{M}/d \ln a$ \cite{Steinhardt}, and the observational data borrowed from \cite{Gong}. In Fig. \ref{fig:f}, we can see that for both models (left panel model 1, right panel model 2) the noninteracting version fits the data better, as said before, especially at the present epoch. This approach was just done to show that interacting versions and $\tilde{c}^{2}$ sharing the same $H(z)$, and so indistinguishable at the background level, can evolve diversely at the perturbative level. However, to definitively discard, or validate, the interacting model, an exact first order perturbation together with the study of matter and radiation power spectrum appears necessary. This will be the subject of a future work.
\section {Conclusions}\label{Conclusions}
As we have seen, at background level, holographic interacting models whose IR cutoff is set by the Hubble's length can be viewed as noninteracting ones whose holographic parameter $\tilde{c}^{2}$ is not constant but varies slowly. Because they share identical Hubble function, they are not distinguishable at the background level. However, this degeneracy can be broken at the perturbative level, since both energy components, i.e., DM and DE, evolve diversely in each scenario. The interacting model and the noninteracting one (section \ref{sec:EX1}) fit the geometrical data rather well.
At the perturbative level, the noninteracting scenario is favored by large scale structure formation. In the second model (Sec. \ref{sec:EX2}), which behaves as a spatially flat $w$CDM model, both interpretations, interacting and $\tilde{c}^{2}(t)$, fit the data well and the values of the parameters seem reasonable. It contains the $\Lambda$CDM as a limiting case but with the coincidence problem alleviated in the interacting scenario. To discriminate between both interpretations, at the perturbative level, the noninteracting scenario appears favored. This should not be surprising, since it mimics rather well the $\Lambda$CDM model.\\ In general, the noninteracting versions seem to be favored by the structure formation; however, the interacting cannot be ruled out just at this point. To go deeper in the matter, a full-fledged perturbative analysis should be undertaken and will be the subject of a future work.
\acknowledgments {We are indebted to Diego Pav\'{o}n for comments, advice and carefully reading the manuscript. We are also indebted to Gaetano Vilasi, Ninfa Radicella and Fernando Atrio-Barandela for fruitful discussions. I.D. and L.P. were supported by an INFN/MICINN collaboration under Grant. No. AIC10-D-000481. This work was partially supported by the ``Ministerio Espa\~{n}ol de Ciencia e innovaci\'{o}n'' under Grant No. FIS2009-13370-C02-01, by the ``Direcci\'{o} de Recerca de la Generalitat de Catalunya'' under Grant No. 2009SGR-00164 and by the Italian Ministero Istruzione Universit`a e Ricerca (MIUR) through the PRIN 2008 grant. I.D. was funded by the ``Univesitat Aut\`{o}noma de Barcelona'' through a PIF fellowship.}
\appendix
\section{Hubble functions considering the radiation component}\label{app:Rad}
To constrain the model with CMB data, it is necessary to take into account the radiation component, since we need to describe the Universe at the last scattering surface, $z_{\star}\approx 1090$, where the said component is no longer negligible. So, expressions like (\ref{eq:H1}) and (\ref{eq:H4}) are not accurate at the last scattering epoch, much less so at earlier times. The presence of radiation invalidates the expression $\rho_{M}=3M_{P}^{2}(1-c^{2})H^{2}$, so to obtain the Hubble function we rewrite the second Friedmann equation as
\be
\label{eq:FE2Rad}
\frac{\dot{H}}{H^{2}} = -\frac{3}{2}\left(1 + w_{X}\Omega_{X} + w_{R}\Omega_{R} \right) \,,
\ee
\no where the subscript $R$ stands for radiation, and the EoS parameter $w_{X}$ does not coincide with the $w$ of Sec. \ref{sec:EX1} and \ref{sec:EX2}. Differentiating Eq.(\ref{eq:rhox}) and using Eqs. (\ref{eq:evolEqIntMX}.2) and (\ref{eq:FE2Rad}), we obtain
\be
\label{eq:wXRad}
w_{X} = -\frac{Q}{(1-c^{2})\rho_{X}H}+\frac{w_{R}}{c^{2}}\left(\frac{H_{0}}{H}\right)^{2}\Omega_{R 0}(1+z)^{4} \,.
\ee
With an expression for the EoS parameter of DE, Eq. (\ref{eq:FE2Rad}) has no analytical solution. However, we can consider two different integration regions, one for $z \leq 50$ and another for $z \geq 50$. In the former, the Universe is DM and DE dominated; in the latter only the higher $z$ terms contribute.
\\\\
In the example considered in Sec. \ref{sec:EX1}, $Q\equiv \Gamma \rho_{X}$, and in the first integration region, $z \leq 50$,  the second term in Eq. (\ref{eq:wXRad}) contributes less than $1\%$, so the Hubble parameter is just as in Eq. (\ref{eq:H1}), and the first Friedmann equation can be approximated by
\be
\label{eq:FE1Rad1}
\left(\frac{H}{H_{0}}\right)^{2}_{z \leq 50} \simeq A^{2}+2A(1-A)(1+z)^{\frac{3}{2}}+(1-A)^{2}(1+z)^{3} \,.
\ee
For the second region, $z \geq 50$,  the first term in Eq. (\ref{eq:wXRad}) contributes no more than $1\%$, so the first Friedman equation is approximated by
\be
\label{eq:FE1Rad2}
\left(\frac{H}{H_{0}}\right)^{2}_{z \geq 50} \simeq (1-A)^{2}(1+z)^{3}+2\Omega_{R 0}(1+z)^{4}\,.
\ee
As the two first terms in Eq. (\ref{eq:FE1Rad1}) are negligible in the region $z \geq 50$, and the second term in Eq. (\ref{eq:FE1Rad2}) is trifling when $z \geq 50$, we can just consider the Hubble function, after redefining $2\Omega_{R 0}\rightarrow \Omega_{R 0}$ as
\be
\label{eq:HRad1}
\left(\frac{H}{H_{0}}\right)^{2} \simeq \left(A+(1-A)^{2}(1+z)^{\frac{3}{2}}\right)^{2}+\Omega_{R 0}(1+z)^{4}\,.
\ee
This Hubble function is the same in both scenarios, the interacting and the $\tilde{c}^{2}$, as when we ignored the radiation component.
\\
Proceeding as before, in the case of the model 2 (Sec. \ref{sec:EX2}) we obtain
\be
\label{eq:HRad2}
\left(\frac{H}{H_{0}}\right)^{2} \simeq \tilde{\Omega}_{M0}(1+z)^{3}+\tilde{\Omega}_{X0}(1+z)^{3-\gamma}+\Omega_{R 0}(1+z)^{4}\, .
\ee
In both cases, $\Omega_{R 0}\approx\Omega_{\gamma 0}(1+0.0227N_{eff})$ as described in \cite{Komatsu}, where $N_{eff}=3.04$ is the effective number of neutrino families.
\section{Lagrangian formulation of the models from sec. \ref{sec:EX1}}\label{app:Lag}
\subsection{Interacting model with 1 scalar field}\label{app:LagA}
In Sec. \ref{sec:Intro}, we have seen that this kind of interacting models, can be described by just one fluid (as long as the evolution of the densities of DE and DM are identical), whose effective EoS parameter is
\be\label{eq:wEffective}
w_{eff} = -A\frac{H_{0}}{H} \quad \text{,}
\ee
\no with $A$ a positive constant. The total effective pressure $P = w_{eff} \rho$ is
\ben\label{eq:PEffective}
P &=& -3^{\frac{1}{2}}A\,M_{P}\,H_{0}\,\rho^{\frac{1}{2}} \label{eq:rhoDePEff} \,,
\een
\no where $P$ and $\rho\equiv\rho_{M}+\rho_{X}$ are the dark sector pressure and energy density from Eqs. (\ref{eq:FE1}) and (\ref{eq:wEffective}). For a standard scalar field $\phi$ minimally coupled to gravity, the action is defined by
\be\label{eq:actionSatndard}
S=\int d^4x\sqrt{-g}\left(\frac{R}{2}+ L_{\phi}(\phi,\chi)\right)
\ee
\no where $\chi=\frac{1}{2}g_{\mu\nu}\partial^{\mu}\phi\partial^{\nu}\phi$ is the kinetic term. Assuming homogeneity and isotropy, the energy (Hamiltonian) density  and the pressure (Lagrangian density) are
\be\label{eq:rhoStandard}
\rho = \frac{1}{2}\dot{\phi}^{2}+V(\phi) \qquad \text{and}\qquad
P = \frac{1}{2}\dot{\phi}^{2}-V(\phi) \,.
\ee
From the last two equations the potential and the kinetic term are
\be\label{eq:dotPhiStandard}
\dot{\phi}^{2}=\rho+P \qquad \text{and}\qquad
V(\phi)=\frac{1}{2}\left(\rho-P\right) \,,
\ee
\no respectively. Using again Eqs.(\ref{eq:FE1}), (\ref{eq:H1}) and (\ref{eq:rhoDePEff}) we obtain $P(z)$ and with (\ref{eq:dotPhiStandard}) we obtain
\be\label{eq:dPhiStandard}
d\phi=\frac{2}{3AH_{0}}\left(\frac{P^{2}}{3A^{2}M_{P}^{2}H_{0}^{2}}+P\right)^{-\frac{1}{2}}dP \, .
\ee
\no After integration we get
\be\label{eq:pDePhiStandard}
L=P=-\frac{3A^{2}M_{P}^{2}H_{0}^{2}}{16}  \left(4e^{-\frac{1}{4}\sqrt{\frac{3}{M_{P}^{2}}}\left(\phi-\phi_{0}\right)}+e^{\frac{1}{4}\sqrt{\frac{3}{M_{P}^{2}}}\left(\phi-\phi_{0}\right)}\right)^2 \, ,
\ee
\no \no where we have defined $\phi_{0}$ as the value of the scalar field at the maximum pressure, i.e., when $z\rightarrow -1$. It is interesting to notice that in the absence of interaction, $A=0$, the pressure of the dark sector is that of standard matter. From (\ref{eq:pDePhiStandard}) and (\ref{eq:dotPhiStandard}) we find the potential
\begin{widetext}
\begin{equation*}\label{eq:VDePhiStandard}
V(\phi)=\frac{3A^{2}M_{P}^{2}H_{0}^{2}}{512} \left(4+e^{\frac{1}{2}\sqrt{\frac{3}{M_{P}^{2}}}\left(\phi-\phi_{0}\right)}\right)^2 \left(16e^{-\sqrt{\frac{3}{M_{P}^{2}}}\left(\phi-\phi_{0}\right)}+24 e^{-\frac{1}{2}\sqrt{\frac{3}{M_{P}^{2}}}\left(\phi-\phi_{0}\right)}+1\right)\, .\end{equation*}
\end{widetext}
So we can describe the whole dark sector by a standard scalar field, that has an effective sound speed $c_{s}^{2}=\frac{\delta{P}}{\delta{\rho}}=1$, even if its adiabatic sound speed $c_{a}^{2}=\frac{\dot{P}}{\dot{\rho}}$ is negative due to the interaction. Had we considered a general k-essence Lagrangian density \cite{ArmendarizMuhanov}, the squared sound speed would have resulted negative.
\subsection{$\tilde{c}^{2}(t)$ model with 1 scalar field}\label{app:LagB}
Here we obtain a Lagrangian density for the DE in the case of the holographic $\tilde{c}^{2}$ model. This is motivated to show that the two models of Sec. \ref{sec:EX1}, described at the background level by the same $H(z)$, differ not only because the energy densities evolve differently but also because their Lagrangians are diverse.\\
To begin with, the DE  density can be expressed as a function of the pressure
\be\label{eq:lastEq1}
\tilde{\rho}=-3A^{2}H_{0}^{2}M_{P}^{2} - 2\tilde{P}\,.
\ee
Proceeding as before we obtain
\ben\label{eq:PhiNoInt}
\tilde{L}&=&-3A^{2}M_{P}^{2}H_{0}^{2} \sec^{2}{\left(\sqrt{\frac{3}{M_{P}^{2}}}\frac{\tilde{\phi}-\tilde{\phi}_{0}}{4}\right)}\\
\text{and}\quad \nonumber\\
\tilde{V}(\tilde{\phi})&=&\frac{3A^{2}M_{P}^{2}H_{0}^{2}}{2}\tan^{2}{\left(\sqrt{\frac{3}{M_{P}^{2}}}\frac{\tilde{\phi}-\tilde{\phi}_{0}}{4}\right)}
\een
As in the previous case, for a k-essence Lagrangian we would have obtained $\tilde{c}_{s}^{2}<0$.

\end{document}